\documentclass{article}
\usepackage{spconf}
\usepackage[binary-units,per-mode=symbol,per-symbol=p]{siunitx}
\usepackage{cite}
\usepackage{adjustbox}
\usepackage[ruled,vlined]{algorithm2e}
\usepackage{amsmath,amssymb,amsfonts}
\usepackage{algorithmic}
\usepackage{graphicx}
\usepackage{textcomp}
\usepackage[colorlinks=false, pdfborder={0 0 0}, breaklinks]{hyperref}
\usepackage{xcolor}
\usepackage{balance}
\usepackage{enumitem}
\setlist{nolistsep,leftmargin=*}
\usepackage{setspace}
\usepackage{wrapfig}
\usepackage{listings}
\usepackage{color}
\usepackage{caption}
\usepackage{subcaption}
\usepackage{todonotes}
\usepackage[longtable]{multirow}

\def\BibTeX{{\rm B\kern-.05em{\sc i\kern-.025em b}\kern-.08em
    T\kern-.1667em\lower.7ex\hbox{E}\kern-.125emX}}

\usepackage{amsmath}

\usepackage{ctable}

\newcommand{\ie}{\emph{i.e.}, }

\begin{document}

\title{Learning Quality from Complexity and Structure: A Feature-Fused XGBoost Model for Video Quality Assessment}

\name{
\begin{tabular}{@{}c@{}}
Amritha Premkumar$^1$ \qquad 
Prajit T Rajendran$^2$ \qquad  
Vignesh V Menon$^3$
\end{tabular}\vspace{-0.5em}
}
\address{}
\address{\small $^1$Department of Computer Science, Rheinland-Pfälzische Technische Universität, Kaiserslautern, Germany \\
\small $^2$CEA, List, F-91120 Palaiseau, Université Paris-Saclay, France\\
\small $^3$Video Communication and Applications Department, Fraunhofer HHI, Berlin, Germany
}

\maketitle

\begin{abstract}
This paper presents a novel approach for reduced-reference \emph{video quality assessment} (VQA), developed as part of the recent VQA Grand Challenge. Our method leverages low-level complexity and structural information from reference and test videos to predict perceptual quality scores. Specifically, we extract spatio-temporal features using Video Complexity Analyzer (VCA) and compute SSIM values from the test video to capture both texture and structural characteristics. These features are aggregated through temporal pooling, and residual features are calculated by comparing the original and distorted feature sets. The combined features are used to train an XGBoost regression model that estimates the overall video quality. The pipeline is fully automated, interpretable, and highly scalable, requiring no deep neural networks or GPU inference. Experimental results on the challenge dataset demonstrate that our proposed method achieves competitive correlation with subjective quality scores while maintaining a low computational footprint. The model’s lightweight design and strong generalization performance suit real-time streaming quality monitoring and adaptive encoding scenarios.
\end{abstract}

\begin{keywords}
Video Quality Assessment (VQA), Video Quality Metrics, Video Complexity Analysis, SSIM, XGBoost Regression.
\end{keywords}

\section{Introduction}
The rapid growth of online video streaming services and user-generated content platforms has led to an exponential increase in video consumption across diverse networks, devices, and screen resolutions~\cite{CiscoForecast}. Ensuring a consistent and satisfactory user experience requires accurate and scalable Video Quality Assessment (VQA) models in such heterogeneous environments. These models must estimate perceived quality in real-time, often under limited computational budgets~\cite{vqtif_ref}.

Traditionally, full-reference (FR) metrics such as Peak Signal-to-Noise Ratio (PSNR)~\cite{psnr_ref1}, Structural Similarity Index (SSIM)~\cite{ssim_ref3}, and Video Multimethod Assessment Fusion (VMAF)~\cite{vmaf_ref1} have been employed for this task. While these metrics are effective in controlled offline evaluations, they require access to undistorted reference videos and are computationally expensive, making them impractical for deployment in real-time streaming applications~\cite{qadra_ref}. To overcome these limitations, recent research has explored machine learning-based VQA methods that learn to predict subjective quality scores from extracted features. While promising, many of these approaches rely on deep neural networks that require large labeled datasets and GPU resources, hindering their deployment in low-latency or resource-constrained environments.

This paper addresses this challenge by proposing a lightweight and interpretable reduced-reference (RR) VQA method that balances predictive performance and computational efficiency. Our approach leverages handcrafted spatio-temporal features derived from the Video Complexity Analyzer (VCA)~\cite{vca_ref}, capturing texture and motion-related complexity in the reference and test videos. We also incorporate structural quality indicators by computing the SSIM on the test video. These features are averaged, and residuals are calculated to quantify the degradation between the reference and the distorted video. The resulting feature vectors are then used to train an XGBoost~\cite{xgboost_ref} regression model that predicts perceived video quality. The proposed pipeline, developed for the VQA Grand Challenge, requires only partial information from the reference video, is fully interpretable, and runs efficiently without the need for neural network inference or specialized hardware. It is thus particularly well-suited for applications such as streaming quality monitoring, adaptive bitrate control, and encoder performance benchmarking.

\emph{Paper outline:} Section~\ref{sec:related} reviews related work in full-, reduced-, and no-reference VQA. Section~\ref{sec:method} details the proposed system architecture and feature extraction pipeline. Section~\ref{sec:evaluation} presents the dataset, training setup, and evaluation results. Finally, Section~\ref{sec:conclusion} concludes the paper and outlines directions for future work.

\section{Related Work}
\label{sec:related}
VQA is a critical task in multimedia systems, aiming to estimate the perceptual quality of video content. Existing VQA approaches are broadly classified into three categories: full-reference (FR), reduced-reference (RR), and no-reference (NR)~\cite{VQ_survey}, depending on the amount of access to the original undistorted video required for quality estimation.

\emph{FR} metrics assume complete availability of the reference video and perform pixel-wise comparisons with the distorted video. Classical examples include PSNR, which measures mean squared error in the pixel domain, and SSIM~\cite{ssim_ref3}, which evaluates structural similarity by modeling local luminance, contrast, and structure. More advanced FR models like MS-SSIM and VMAF~\cite{vmaf_ref1} combine multiple quality cues including motion, texture, and perceptual features~\cite{chen_yixu_ref}. Although FR methods are widely adopted for offline benchmarking and encoder evaluations, they are impractical in real-time streaming applications due to the need for access to the full reference video and high computational complexity.

\emph{RR} methods attempt to balance practicality and accuracy by extracting a compact set of features from the reference video that can be transmitted alongside the distorted content. Early RR methods include RR-PSNR~\cite{rr_psnr1}, RR-SSIM~\cite{rr_ssim1}, and STRRED~\cite{rr_strred1}, which transmit signal statistics or multi-scale coefficients. Compared to FR methods, these approaches significantly reduce the bandwidth required for quality estimation. However, most traditional RR methods focus on handcrafted signal-level features and lack adaptability to diverse content and distortion types.

\emph{NR or blind VQA} methods perform quality prediction without access to the reference video~\cite{p1203_ref}. Early NR models such as NIQE and BRISQUE use natural scene statistics (NSS) to model deviations from expected patterns in high-quality content. While computationally efficient, their performance degrades under complex distortion types. Recent advances in deep learning have led to data-driven NR models like~\cite{funque} that utilize CNNs or Transformers trained on large-scale datasets like KoNViD-1k~\cite{hosu2017konstanz} and YouTube-UGC~\cite{ugc-ref}. These models capture semantic and spatio-temporal patterns and often outperform traditional handcrafted approaches. However, they are data-hungry, less interpretable, and computationally intensive—posing challenges for real-time deployment in bandwidth-constrained or embedded systems.

Our work sits in the RR category, bridging the gap between classical handcrafted methods and black-box deep learning models. We utilize perceptually meaningful, domain-specific handcrafted features extracted via VCA~\cite{vca_ref}, which capture spatial texture and temporal dynamics. Additionally, we integrate SSIM-based structural features from the test video to encode visual distortions. These features are fused and fed into a lightweight XGBoost regression model, offering both interpretability and efficient training on modest data sizes. In contrast to deep learning-based NR models, our approach requires only partial reference-side information. It can be deployed in scenarios where full-reference data is unavailable, but limited metadata, such as streaming pipelines or encoder testing frameworks, can be shared.

\begin{figure*}[t]
\centering
\resizebox{0.99\textwidth}{!}{
\begin{tikzpicture}[
    node distance=1.4cm and 1.5cm,
    every node/.style={draw, minimum height=1cm, align=center, font=\small},
    io/.style={minimum width=2.8cm},
    process/.style={minimum width=3.6cm},
    >=latex
]

\node[io] (ref) {Reference Video};
\node[io, below of= ref] (test) {Test Video};

\node[process, right of= ref, xshift = 6em] (vca_ref) {VCA Feature\\Extraction (Ref)};
\node[process, right of= test, xshift = 6em] (vca_test) {VCA Feature\\Extraction (Test)};

\node[process, right of= vca_ref, xshift = 8em] (avg_ref) {Feature\\Averaging (Ref)};
\node[process, right of= vca_test, xshift = 8em] (avg_test) {Feature\\Averaging (Test)};

\node[process, right of= avg_ref, yshift=-1.0cm, xshift = 8em] (resid) {Residual\\Computation};

\node[process, below of= resid, yshift=-0.0cm] (ssim) {SSIM (Ref vs Test)};

\node[process, right of= resid, xshift=4.5cm] (xgb) {Regression Model:\\XGBoost};

\node[process, right of= xgb, xshift=8em] (score) {VQA score};

\draw[->] (ref) -- (vca_ref);
\draw[->] (test) -- (vca_test);

\draw[->] (vca_ref) -- (avg_ref);
\draw[->] (vca_test) -- (avg_test);

\draw[->] (avg_ref.east) -- ++(0.25,0) |- (resid.west);
\draw[->] (avg_test.east) -- ++(0.25,0) |- (resid.west);

\draw[->] (resid) -- (xgb);

\draw[->] (xgb) -- (score);

\draw[->] (ref.west) -| ++(-0.4,-3.5) -| (ssim.south);
\draw[->] (test) |- ++(0,-1.0) -- (ssim.west);

\draw[->] (ssim.east) -- ++(1.0,0) -| (xgb.south);

\end{tikzpicture}}
\caption{Architecture of the proposed VQA system.}
\label{fig:vqa_pipeline}
\end{figure*}
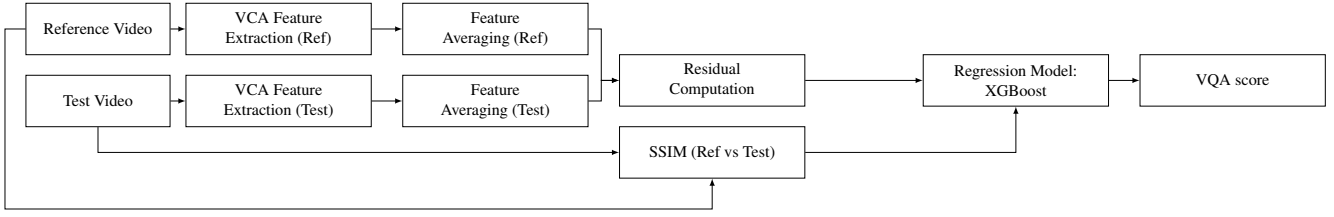

\section{Proposed Method}
\label{sec:method}
In this section, we present our proposed reduced-reference video quality assessment (VQA) framework, grounded in principles from information theory. The method is composed of five stages: (1) VCA-based feature extraction, (2) temporal feature averaging, (3) residual computation, (4) structural quality modeling via SSIM, and (5) nonlinear quality regression using XGBoost as shown in Fig.~\ref{fig:vqa_pipeline}. We frame each component through the lens of mutual information theory, drawing connections between the empirical feature representations and theoretical concepts of information preservation and distortion.

\subsection{Information-Theoretic Formulation of VQA}
Let $X$ denote the source (reference) video and $\hat{X}$ the distorted version produced through lossy compression and transmission. From an information-theoretic viewpoint, this transformation constitutes a noisy channel, with video quality $Y$ as the perceptual reconstruction fidelity interpreted by a human or machine observer~\cite{wang_mse}. Ideally, perceived quality should be a monotonic function of the mutual information between $X$ and $\hat{X}$:

\begin{equation}
    Y \propto I(X; \hat{X})
\end{equation}

Direct computation of $I(X; \hat{X})$ is infeasible due to the high dimensionality of video data. Instead, we introduce handcrafted proxy features that capture relevant spatial, temporal, and structural distortions between $X$ and $\hat{X}$ and use machine learning to infer $Y$.

\begin{table}[t]
\centering
\caption{Interpretation of model components in information theory}
\resizebox{0.8\columnwidth}{!}{
\begin{tabular}{l|l}
\toprule
\toprule
\textbf{Component} & \textbf{Information-Theoretic Role} \\
\midrule
\midrule
$\boldsymbol{x}, \hat{\boldsymbol{x}}$ & Empirical features of source/channel output \\
$\bar{\boldsymbol{x}}, \bar{\hat{\boldsymbol{x}}}$ & Marginal feature distributions \\
$\boldsymbol{r}$ & Proxy for KL divergence / information loss \\
$\mu_{\text{SSIM}}$ & Structural mutual information estimate \\
$f(\cdot)$ & Empirical MI-to-quality mapping \\
\bottomrule
\bottomrule
\end{tabular}
}
\end{table}

\subsection{VCA Feature Extraction as Mutual Information Embedding}
We extract low-level features from reference and distorted videos using the Video Complexity Analyzer (VCA)~\cite{vca_ref, jtps_ref}. Each frame $i$ yields a 7-dimensional feature vector capturing DCT energy-based texture and structure statistics:

\begin{equation}
    \boldsymbol{x}_i = [E_{Y,i}, h_i, L_{Y,i}, E_{U,i}, L_{U,i}, E_{V,i}, L_{V,i}]
\end{equation}

\begin{equation}
    \hat{\boldsymbol{x}}_i = [\hat{E}_{Y,i}, \hat{h}_i, \hat{L}_{Y,i}, \hat{E}_{U,i}, \hat{L}_{U,i}, \hat{E}_{V,i}, \hat{L}_{V,i}]
\end{equation}

Here, $E_Y$ is the average texture energy in the luma channel, $h$ is the gradient magnitude, $L_Y$ is the average luminance, and the remaining features describe chroma complexity. These handcrafted features serve as compact empirical embeddings of the source and channel output, enabling an approximate estimation of the mutual information lost during compression~\cite{mittal_ref}.

\subsection{Feature Averaging and Temporal Pooling}
For each video sequence comprising $N$ frames, we apply temporal pooling to obtain segment-level representations:

\begin{equation}
    \bar{\boldsymbol{x}} = \frac{1}{N} \sum_{i=1}^{N} \boldsymbol{x}_i, \quad \bar{\hat{\boldsymbol{x}}} = \frac{1}{N} \sum_{i=1}^{N} \hat{\boldsymbol{x}}_i
\end{equation}

The averaged vectors $\bar{\boldsymbol{x}}$ and $\bar{\hat{\boldsymbol{x}}}$ summarize the statistical complexity over the entire video segment for both reference and distorted streams. They estimate the empirical marginal distributions $P_X$ and $P_{\hat{X}}$ over the feature space in information-theoretic terms.

\subsection{Residual Computation as Information Loss Proxy}
To model distortion, we compute residuals between the reference and distorted feature means:

\begin{equation}
    \boldsymbol{r} = \bar{\boldsymbol{x}} - \bar{\hat{\boldsymbol{x}}}
\end{equation}

Each component $r_j = \bar{x}_j - \bar{\hat{x}}_j$ approximates the information lost in the $j$-th feature dimension. Assuming Gaussian feature distributions, this difference relates to the Kullback–Leibler divergence between $P_X$ and $P_{\hat{X}}$, \ie

\begin{equation}
    D_{\text{KL}}(P_X || P_{\hat{X}}) \approx \frac{1}{2} \|\boldsymbol{r}\|_2^2
\end{equation}

Thus, $\boldsymbol{r}$ is a proxy for information degradation due to lossy encoding. The larger the magnitude of residuals, the greater the perceived quality loss.

\subsection{Structural Similarity as Mutual Information Estimate}
While complexity-based residuals capture energy and texture losses, perceptual quality also depends on structural distortions. We incorporate Structural Similarity Index (SSIM) as a perceptual fidelity measure. Let $\text{SSIM}_i$ be the similarity between reference and distorted frames at time $i$:

\begin{equation}
    \mu_{\text{SSIM}} = \frac{1}{N-1} \sum_{i=1}^{N-1} \text{SSIM}_i
\end{equation}

SSIM has been shown to approximate mutual information in the structural domain~\cite{sheikh_ref}. Specifically, for certain Gaussian assumptions, SSIM is a lower bound on the normalized mutual information:

\begin{equation}
    \text{SSIM}(X, \hat{X}) \approx \frac{I(X; \hat{X})}{H(X)}
\end{equation}

Hence, $\mu_{\text{SSIM}}$ provides complementary information to $\boldsymbol{r}$, offering insight into contrast, luminance consistency, and perceptual fidelity beyond raw energy loss.

\subsection{Feature Fusion and Regression via XGBoost}

The fused feature vector is constructed as:

\begin{equation}
    \boldsymbol{z} = [\boldsymbol{r} \mid \mu_{\text{SSIM}}] \in \mathbb{R}^8
\end{equation}

This vector forms a compressed representation of the approximate information difference between $X$ and $\hat{X}$ across both complexity and structure domains. To map this vector to a predicted perceptual quality score, we use a regression function $f(\cdot)$ implemented via XGBoost~\cite{xgboost_ref}:

\begin{equation}
    \hat{y} = f(\boldsymbol{z})
\end{equation}

XGBoost is a scalable ensemble learning algorithm based on gradient-boosted decision trees. It is particularly well-suited for tabular data with heterogeneous features, low dimensionality, and non-linear interactions. Unlike deep networks, it offers fast training, interpretability, and generalization with modest data requirements.

From an information-theoretic angle, $f(\cdot)$ can be viewed as an empirical estimator of the mapping:

\begin{equation}
    f: \text{MI-approximation} \rightarrow \text{Subjective Quality}
\end{equation}


\section{Evaluation}
\label{sec:evaluation}
\subsection{Dataset}
The Image and Video Engineering (LIVE) lab at the University of Texas at Austin (sponsored by Amazon Prime Video) created the dataset provided for this challenge. It contained 54 pristine, high-quality source videos. These videos include 31 open-source videos from the 8K HDR AVT-VQDB-UHD-2-HDR dataset, 10 Video on Demand (VoD) videos, and 10 Live Sports videos from Amazon Prime Video’s internal source, and three anchor videos from the LIVE HDR or SDR database. All videos are in the BT.2020 color gamut and are quantized using the PQ Optical-Electronic Transfer Function (OETF). Each video sequence had a duration of approximately 7 seconds and included static metadata of the HDR10 standard~\cite{chen2025hdrsdrvqa}.

In this challenge, we only utilized the 31 open-sourced contents—20 open-sourced videos, along with their processed video sequences (PVS), will be provided for training, while the remaining videos and their PVS were reserved for testing. The training videos comprise 360 videos (180 HDR and 180 SDR), while the testing videos include 198 videos (99 HDR and 99 SDR). The videos with 3840x2160 and 50Mbps ladder were used as the reference videos.

\subsection{Hyperparameter Tuning}
We optimized the XGBoost regressor using Optuna, a Bayesian hyperparameter search framework~\cite{akiba2019optuna}. A total of 50 trials were performed, exploring the search space: \texttt{n\_estimators} $\in$ [50, 300],  \texttt{max\_depth} $\in$ [3, 10], \texttt{learning\_rate} $\in$ [0.01, 0.3] (log scale),  \texttt{subsample} $\in$ [0.6, 1.0], and \texttt{colsample\_bytree} $\in$ [0.6, 1.0],

The best configuration identified by Optuna achieved the highest Pearson correlation coefficient (PCC) on the validation set and is \texttt{n\_estimators:} 95, \texttt{max\_depth:} 8, \texttt{learning\_rate:} 0.072, \texttt{subsample:} 0.999, and \texttt{colsample\_bytree:} 0.852.

\subsection{Relevance of features}
The importance of the input features to the XGBoost-based fusion model is analyzed using the SHAP features~\cite{feature_imp}. It is visualized in Fig.~\ref{fig:rrtif_feat_imp}. It is observed that the $r_{\text{E}}$ feature contributes the most to the quality estimation.
\subsection{Accuracy}
\begin{figure}[t]
\centering
\begin{subfigure}{0.56\columnwidth}
    \centering
    \includegraphics[width=\textwidth]{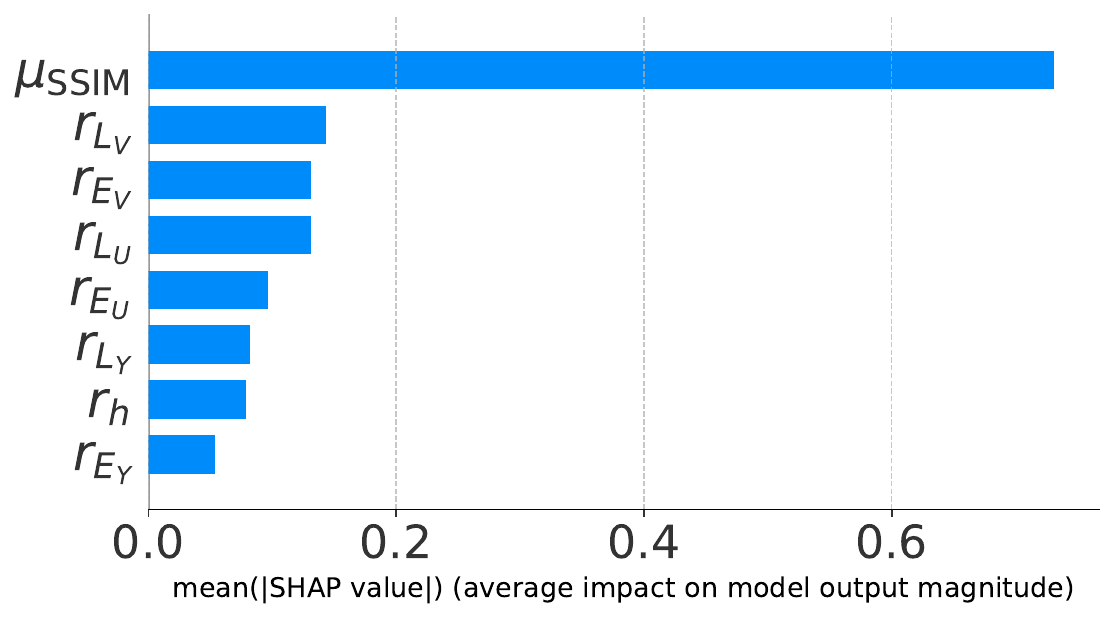}
    \caption{Feature importance of the XGBoost model.}
        \label{fig:rrtif_feat_imp}
\end{subfigure}
\hfill
\begin{subfigure}{0.43\columnwidth}
    \centering
    \includegraphics[width=\textwidth]{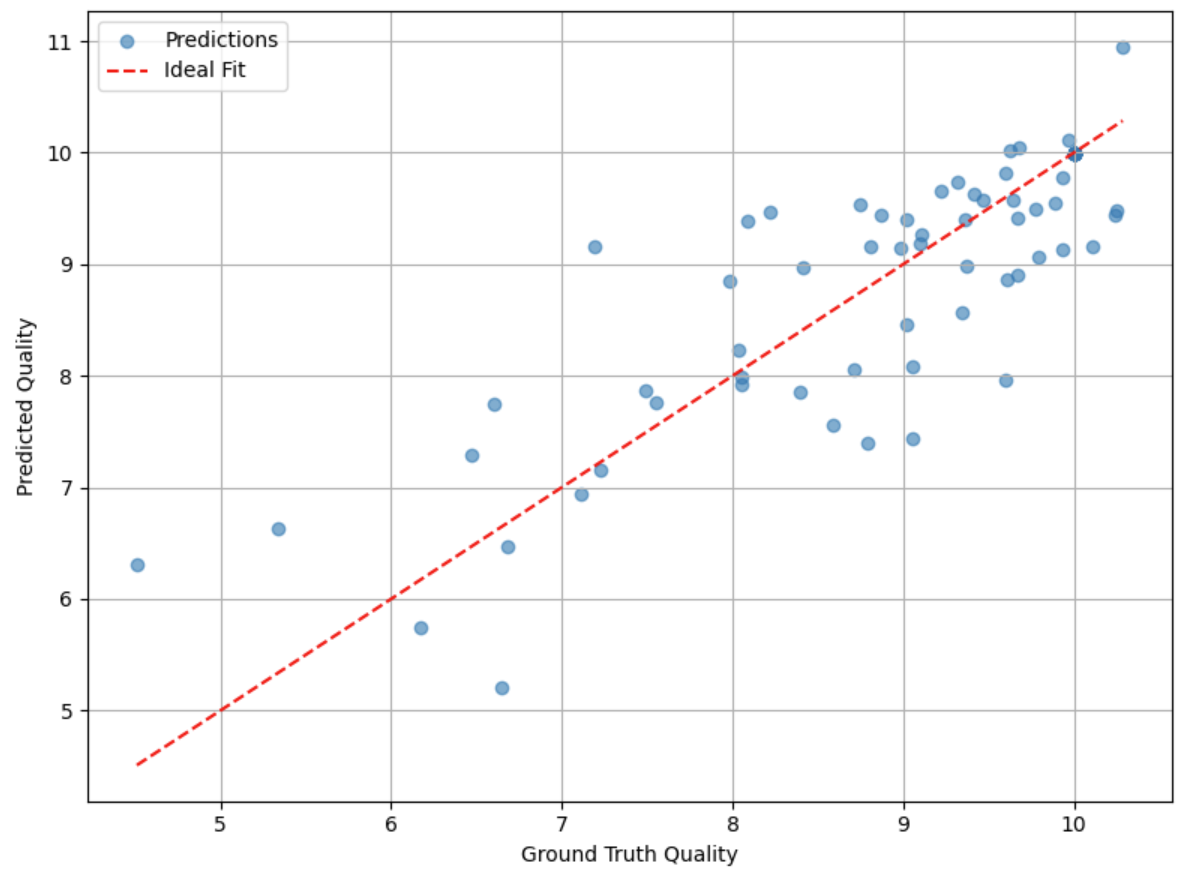}
    \caption{Scatterplot of our VQA scores and the ground truth.}
        \label{fig:rrtif_scatter_plot}
\end{subfigure}
\caption{Prediction results.}
\label{fig:rrtif_time_energy_plot}
\end{figure}

To evaluate the effectiveness of our proposed method, we compute the Spearman Rank Order Correlation Coefficient (SROCC), Pearson Linear Correlation Coefficient (PLCC)~\cite{pcc_ref}, Kendall Rank Order Correlation Coefficient (KROCC), and Root Mean Square Error (RMSE) between the predicted quality scores and the ground truth subjective scores. Our model achieves an average PLCC of \SI{0.787}{}, indicating a strong correlation between the predicted and actual perceptual quality. As shown in Fig.~\ref{fig:rrtif_scatter_plot}, the scatterplot reveals tight clustering around the diagonal, confirming the reliability of our predictions. Furthermore, Table~\ref{fig:rrtif_time_energy_plot} presents a quantitative comparison against widely used metrics such as P1204\_3, VMAF~\cite{vmaf_ref1}, and PSNR-Y~\cite{psnr_ref1}. While P1204\_3 and VMAF slightly outperform our method in terms of PCC and RMSE, our approach still performs significantly better than PSNR-Y, especially in correlation with human perception. This highlights the potential of our lightweight, interpretable model in scenarios where computational efficiency is critical, such as real-time streaming and adaptive encoding.

\begin{table}[t]
\caption{Performance on the test set.}
\centering
\resizebox{0.90\columnwidth}{!}{
\begin{tabular}{l||c|c|c|c}
\specialrule{.12em}{.05em}{.05em}
\specialrule{.12em}{.05em}{.05em}
Metrics & SROCC & PLCC & KROCC & RMSE \\
\specialrule{.12em}{.05em}{.05em}
\specialrule{.12em}{.05em}{.05em}
P1204\_3 & 0.925 & 0.935 & 0.762 & 0.496 \\
VMAF~\cite{vmaf_ref1} & 0.905 & 0.901 & 0.725 & 0.597 \\
PSNR-Y~\cite{psnr_ref1} & 0.674 & 0.677 & 0.48 & 1.015\\
\hline
Our method & 0.832 & 0.787 & 0.632 & 0.862\\
\specialrule{.12em}{.05em}{.05em}
\specialrule{.12em}{.05em}{.05em}
\end{tabular}}
\end{table}

\subsection{Computational Time}
\begin{figure}[t]
\centering
\includegraphics[width=0.7\columnwidth]{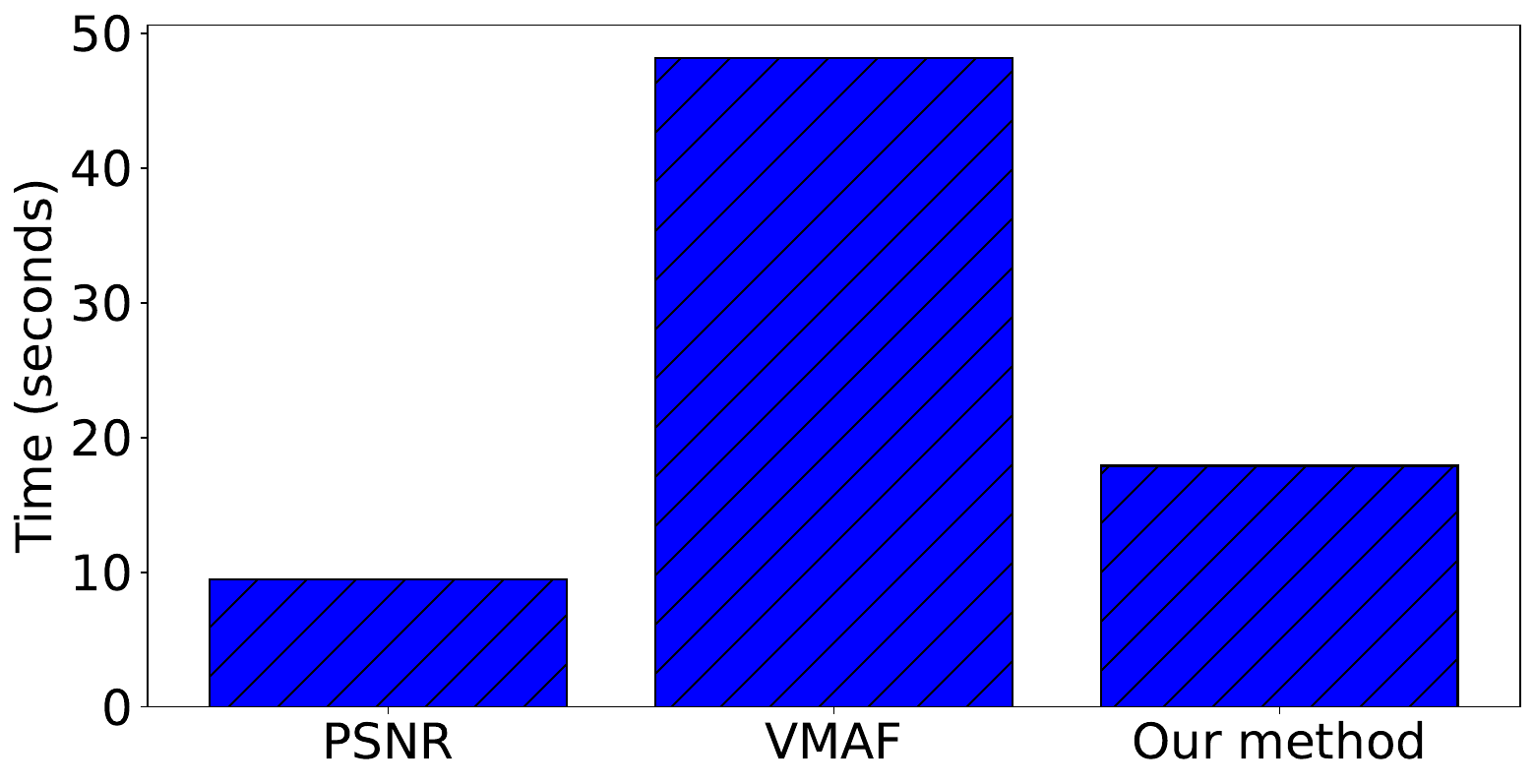}
\caption{Comparative analysis of computational time.}
\label{fig:time_comp}
\end{figure}

We compare the runtime performance of our method against PSNR and VMAF to evaluate its suitability for real-time applications. As shown in Fig.~\ref{fig:time_comp}, our method demonstrates significantly lower computational time than VMAF, with execution times closer to PSNR. Specifically, the average runtime of our model is one-third of the VMAF computational time. This substantial efficiency gain, achieved without reliance on deep learning or GPU acceleration, underscores the practicality of our lightweight feature-based approach. The reduced computational footprint makes it well-suited for integration into streaming pipelines and edge deployments.

\subsection{Limitations}
While the proposed system demonstrates competitive prediction accuracy and offers practical benefits regarding interpretability and efficiency, it has limitations. First, the model operates under a reduced-reference setting, requiring access to complexity features from the reference video. Although this is lighter than full-reference metrics, it still limits the applicability in truly blind scenarios where reference information is entirely unavailable. Second, based on DCT energy and SSIM statistics, the handcrafted feature design may not fully capture semantic or perceptual degradations in highly dynamic or content-rich scenes. Unlike deep learning models that automatically learn multi-level abstractions, the current model relies on fixed features that may not generalize well to unseen distortion types or novel content genres. Future extensions could explore integrating lightweight motion features or transformer-based temporal attention modules.

\section{Conclusion}
\label{sec:conclusion}
This paper presented a lightweight, interpretable RR-VQA framework developed for the VQA Grand Challenge. The proposed method fuses handcrafted low-level complexity features extracted via VCA with structural quality cues obtained from SSIM statistics. By computing residuals between reference and distorted video features, the model effectively captures both signal-level degradation and perceptual structural changes. These fused features are used to train an XGBoost regressor, enabling efficient and accurate quality prediction without the need for deep learning or GPU acceleration. Our method achieves a Pearson correlation coefficient of \SI{0.787}{} on the test set, outperforming classical metrics like PSNR and approaching the performance of more complex methods such as VMAF and P.1204\_3. Notably, the system requires only partial reference-side information, making it ideal for deployment in practical scenarios such as streaming quality monitoring, adaptive encoding, and encoder benchmarking, where full-reference data may not be accessible.

In future work, we aim to enhance temporal modeling by incorporating attention-based architectures such as temporal transformers to capture long-range dependencies. Additionally, we plan to extend the method to support a wider variety of distortion types and content domains through domain adaptation, and further optimize the system for real-time inference in streaming clients and edge computing environments.

\setlength{\parskip}{0pt}
\setlength{\itemsep}{0pt}
 
\bibliographystyle{IEEEtran}
{\linespread{0.4}\selectfont\bibliography{references.bib}}
\balance
\end{document}